\documentclass[11pt,a4paper]{article}
\usepackage{jheppub}
\usepackage{amsmath}
\usepackage[most]{tcolorbox}
\usepackage{dsfont}
\usepackage{ulem}
\usepackage{natbib}
\usepackage{xcolor}
\usepackage[hang,flushmargin]{footmisc}
\usepackage{tikz-cd}
\usepackage{enumitem}
\usepackage{amsfonts,amssymb, amscd,amsmath,latexsym,amsbsy,bm}
\usepackage{stmaryrd}
\usepackage{todonotes}
\usepackage{stackrel}
\usepackage{amsthm}

\usepackage{graphicx}

\title{Notes on the lens integral pentagon identity}

\author{H. Kubra Bag$^{a}$, Osman Ergec$^{a}$, and Ilmar Gahramanov$^{a,b,c}$}
\affiliation{
	$^a$ {Department of Physics, Bogazici University, 34342 Bebek, Istanbul, Turkey}\\[-0.5cm]
	
	$^{b}$ Steklov Mathematical Institute of Russ. Acad. Sci., Gubkina str. 8, 119991 Moscow, Russia\\[-0.5cm]
	
	$^{c}$ Department of Mathematics, Khazar University,  Mehseti St. 41, AZ1096, Baku, Azerbaijan \\
}
\vspace{0.2cm}

\emailAdd{hatice.bag@boun.edu.tr}
\emailAdd{osman.ergec@boun.edu.tr}
\emailAdd{ilmar.gahramanov@boun.edu.tr}

\abstract{We obtain the lens integral pentagon identity for three-dimensional mirror dual theories in terms of hyperbolic hypergeometric functions via reduction of equality for $S_b^3/\mathbb Z_r$ $\mathcal N=2$ supersymmetric partition functions of a certain supersymmetric IR duality. }


\begin{document}
	\maketitle
	\flushbottom

\section{Introduction}

Pentagon relations have plenty of important applications in the theory of integrable lattice models of statistical mechanics, topological field theories, knot invariants, etc  \cite{Kashaev:1996rz,Kashaev:2014rea,Bozkurt:2018xno,Bozkurt:2020gyy,Allman:2017xma,Gahramanov:2014ona,Gahramanov:2013rda,Gahramanov:2016wxi,Gahramanov:2022jxz,Gahramanov:2021pgu,Dede:2022ofo,Kashaev:2012cz,colazzo2020set}. From an algebraic perspective, pentagon relations are related to the Heisenberg double \cite{kashaev1997heisenberg,militaru2004heisenberg,aghaei2019heisenberg}. In this paper, we obtain an integral pentagon relation arising from supersymmetric gauge theory computations. In order to obtain new relation, we consider a certain three-dimensional $\mathcal N=2$ supersymmetric duality on lens space $S_b^3/\mathbb Z_r$. By taking several limits of the partition functions for supersymmetric theories we end up with a new pentagon identity for hyperbolic hypergeometric functions.

\section{Notations}

In this paper we use the so-called hyperbolic gamma function which has the following integral representation:
\begin{equation}
    \gamma ^{(2)}(y; \omega_1 , \omega_2) = \exp{\bigg[ - \int_{0}^{\infty} \bigg( \frac{\sinh{(2y - \omega_1 - \omega_2)}x}{2\sinh{(\omega_1 x)}\sinh{(\omega_2 x)}} - \frac{2y-\omega_1 - \omega_2}{2\omega_1\omega_2 x} \bigg) \frac{dx}{x} \bigg]},
\end{equation}
The $\gamma^{(2)}(y;\omega_1, \omega_2)$ function has following asymptotics:
\begin{equation}
    \begin{split}
    \lim_{y\to \infty} e^{\frac{i\pi}{2}B_{2,2}(y;\omega_1,\omega_2)}\gamma^{(2)}(y;\omega_1,\omega_2)=1, \arg{\omega_1} <\arg{y} < \arg{\omega_2} + \pi,
    \\
    \lim_{y\to \infty} e^{-\frac{i\pi}{2}B_{2,2}(y;\omega_1,\omega_2)}\gamma^{(2)}(y;\omega_1,\omega_2)=1, \arg{\omega_1} - \pi <\arg{y} < \arg{\omega_2},
    \end{split}\label{eq:asymptotics}
\end{equation}
where the second order Bernoulli polynomial $B_{2,2}(y;\omega_1,\omega_2)$ is defined as follows
\begin{equation}
    B_{2,2}(y;\omega_1,\omega_2) = \frac{y^2}{\omega_1\omega_2} - \frac{y}{\omega_1} - \frac{y}{\omega_2} + \frac{1}{6}\bigg(\frac{\omega_1}{\omega_2} + \frac{\omega_2}{\omega_1} \bigg) +\frac{1}{2} .
\end{equation}

\section{Lens partition function}

In \cite{Imamura:2012rq,Imamura:2013qxa}, the lens partition function was derived via the supersymmetric localization technique, and in \cite{Benini:2011nc,Yamazaki:2013fva,Eren:2019ibl,Gahramanov:2016ilb}, authors used the dimensional reduction of the four-dimensional lens superconformal index. Here we briefly outline some basic ingredients of $\mathcal N=2$ $S_b^3/\mathbb Z_r$ supersymmetric partition function. We mostly follow the notations of \cite{Catak:2021coz,Bozkurt:2020gyy}. A reader interested in details is encouraged to look at \cite{Imamura:2012rq,Imamura:2013qxa,Nieri:2015yia,Gahramanov:2016ilb}. 

By making the identification $(x,y)\sim(e^{\frac{2\pi i}{r}}x,e^{\frac{2\pi i}{r}}y)$ one can obtain the lens space $S_b^3/\mathbb Z_r$ from the squashed three sphere 
\begin{equation}
    S_b^3= \{(x,y)\in\mathbb{C}^2, b^2|x|^2+b^{-2}|y|^2=1\} \;.
\end{equation}
The supersymmetric partition function on this manifold can be reduced to the following matrix model via supersymmetric localization
\begin{equation}
    Z=\sum_m\int \frac{1}{|W|}\prod_j^{rank G} \frac{dz_j}{2\pi i r}Z_{\text{classical}}[z,m]Z_{\text{one-loop}}[z,m] \;,
\end{equation}
where the sum is over the holonomies $m=\frac{r}{2\pi}\int_C A_\mu dx^\mu$, $C$ is the non-trivial cycle on $S^3_b/\mathbb{Z}_r$ and $A_\mu$ is the gauge field, and the integral is over the Cartan subalgebra of the gauge group and $z_j$ variables stand for the gauge fugacities. The prefactor $|W|$ corresponds to the order of the gauge group, which is broken by the holonomy into a product of $r$ subgroups. 

The classical term $Z_{\text{classical}}$ includes the contributions coming from the classical action of the Chern-Simons term and Fayet-Iliopoulos term.

The one-loop contribution of chiral multiplets is given in terms of hyperbolic hypergeometric function. A chiral multiplet in the fundamental representation of the gauge group contributes as 
\begin{equation}
    \begin{split}
        Z_{\rm chiral} =
\prod_{\rho}\prod_{\phi} & e^{\frac{i\pi}{2r} ((\rho(m) + \phi(n)) (r-\rho(m) - \phi(n))  - (r-1)(\rho(m)+\phi(n))^2 )} \\
& \times\gamma^2(i(\rho(z) + \phi(\Phi)) + \omega_1 (\rho(m) + \phi(n)) + (\omega_1 + \omega_2)/2; \omega_1 r, \omega_1 + \omega_2) \\
& \times \gamma^2(iz + \omega_2(r-(\rho(m) - \phi(n) ) + (\omega_1 + \omega_2)/2 ; \omega_2 r, \omega_1 + \omega_2) .
    \end{split}
\end{equation}
where $\rho$ and $\phi$ weights of the representation of the gauge group, and the flavor group, respectively. Note that the Weyl weight of a chiral multiple is absorbed into the flavor fugacity. The one-loop contribution of the vector multiplet in our case is trivial since we consider only $U(1)$ gauge theories. 

The partition function is completely determined by the group-theoretical data of the theory. Therefore once we know the group-theoretical data of a theory on $S_b^3/\mathbb Z_r$, we can write down the partition function in terms of hyperbolic hypergeometric integral.

\section{Supersymmetric duality}

Our starting point is the following supersymmetric three-dimensional $\mathcal N=2$ IR duality:

\begin{itemize}

\item \textbf{Theory A:} Theory with $U(1)$ gauge symmetry and $SU(3)_L \times SU(3)_R$ flavor group, having three fundamental chiral multiples and three anti-fundamental chiral multiplets and no Chern-Simons term.

\item \textbf{Theory B:} Theory  without gauge degrees of freedom having the same global symmetries. There are  nine ``mesons'', transforming in the fundamental representation of the flavor group $SU(3)_L \times SU(3)_R$.

\end{itemize}

Partition functions of dual theories agree and as a result one obtains the following hyperbolic hypergeometric integral identity \cite{Sarkissian:2018ppc}
\begin{align}
        \sum\limits_{y = 0}^{[r/2]} \epsilon(y)  \int_{-i\infty}^{i\infty} \nonumber
        \prod _{i=1}^{3}\gamma^{(2)}(a_i -z +\omega_1(u_i - y); \omega_1 r, \omega_1 + \omega_2) \gamma^{(2)}(a_i - z +\omega_2(r-(u_i - y)); \omega_2 r, \omega_1 + \omega_2)\\ \nonumber
        \times \gamma^{(2)}( b_i + z + \omega_1(v_i + y); \omega_1 r, \omega_1 + \omega_2) \gamma^{(2)}(b_i + z + \omega_2(r-(v_i + y)); \omega_2 r, \omega_1 + \omega_2) \frac{dz}{ir\sqrt{\omega_1\omega_2}}\\ 
        = \prod\limits_{i,j=1}^{3} \gamma^{(2)}(a_i+ b_j +\omega_1(u_i + v_j); \omega_1 r, \omega_1 + \omega_2) \gamma^{(2)}(a_i + b_j + \omega_2(r-(u_i + v_j)); \omega_2 r, \omega_1 + \omega_2)\label{eq:bailey-limit}
\end{align}
with the balancing conditions $\sum_i a_i+b_i=\omega_1+\omega_2$ and $\sum_i u_i+v_i=r$. Here $a_i$ are $SU(3)_L$ flavor fugacities and $b_i$ are $SU(3)_R$ flavor fugacities. The $\epsilon(y)$ function is defined as $\epsilon(0)=\epsilon(\lfloor\frac{r}{2}\rfloor)=1$ and $\epsilon(y)=2$ otherwise.

Note that this integral identity can be written as the star-triangle relation and pentagon identity \cite{Bozkurt:2020gyy,Mullahasanoglu:2021xyf,Sarkissian:2018ppc}. The Bailey pair for this identity was constructed in \cite{Gahramanov:2022jxz} (see, also \cite{Gahramanov:2015cva}).  A similar identity was discussed for the $S^3$ sphere partition functions (without summation) in \cite{Spiridonov:2010em,Kashaev:2012cz} (see also \cite{Kels:2013ola}) and for the $S^2 \times S^1$ partition function in \cite{Gahramanov:2013rda,Gahramanov:2014ona,Gahramanov:2016wxi}.

Another supersymmetric duality considered in this paper is mirror symmetry. Three-dimensional mirror symmetry was introduced in \cite{Intriligator:1996ex} for $\mathcal N = 4$ supersymmetric gauge theories, and   in \cite{Aharony:1997bx} for $\mathcal N=2$ gauge theories. The simplest example of $\mathcal N = 2$ mirror symmetry is the duality between supersymmetric quantum electrodynamics with one flavor and the free Wess-Zumino theory:

\begin{itemize}

\item \textbf{Theory A:} Theory has one flavor consisting of two chiral fields and one vector multiplet. This theory possesses two U$(1)$ global symmetries: one is the topological U$(1)_{J}$, and the other is the flavor symmetry U$(1)_{A}$. 

\item \textbf{Theory B:} The mirror theory, the free Wess-Zumino model (this theory is often called the XYZ model) is the theory containing three chiral fields interacting through the trilinear superpotential. This theory has also two U$(1)$ global symmetries, named U$(1)_{V}$ and U$(1)_{A}$.

\end{itemize}

In the context of mirror symmetry, one  identifies U$(1)_{J}$ and U$(1)_{A}$ of the SQED theory with U$(1)_{V}$ and U$(1)_{A}$ of the corresponding Wess-Zumino model. The equality of lens partition functions for mirror dual theories can be written as follows
\begin{align} \nonumber
        \sum\limits_{y = 0}^{[r/2]} \epsilon(y)  \int_{-i\infty}^{i\infty} &e^{\frac{2\pi i}{\omega_1 \omega_2 r}(z + \alpha) (\frac{3\alpha + \beta - \omega}{2}) + \frac{i \pi}{r}(y + n_{\alpha}) (2n_{\beta} - r)}
        \gamma^{(2)}(z + \alpha +\omega_1(y+n_{\alpha}); \omega_1 r, \omega)\\ \nonumber
        &\times \gamma^{(2)}(z + \alpha +\omega_2(r-y-n_{\alpha})); \omega_2 r, \omega)\gamma^{(2)}(-z+\alpha +\omega_1(-y + n_{\alpha}); \omega_1 r, \omega)\\ \nonumber
        &\times \gamma^{(2)}(-z+\alpha +\omega_2(r + y -n_{\alpha})); \omega_2 r, \omega)\frac{dz}{ir\sqrt{\omega_1\omega_2}} \\ \nonumber
        &=e^{\frac{2\pi i }{\omega_1 \omega_2 r} (\alpha) (\frac{2\beta - \omega}{2}) + \frac{i \pi}{r} ( -n_{\alpha})(2n_{\beta} - r)} 
        \gamma^{(2)}(\omega - \alpha - \beta + \omega_1(r - n_{\alpha} - n_{\beta}); \omega_1 r, \omega)\\ \nonumber
        &\times \gamma^{(2)}(\omega - \alpha - \beta + \omega_2(n_{\alpha} + n_{\beta});\omega_2 r, \omega)
        \gamma^{(2)}(\beta - \alpha + \omega_1 (n_{\beta} - n_{\alpha}); \omega_1 r, \omega) \\ \nonumber
        & \times \gamma^{(2)}( \beta - \alpha + \omega_2(r-n_{\beta}+n_{\alpha});\omega_2 r, \omega)\\ 
        & \times \gamma^{(2)}(2\alpha + \omega_1 (2n_{\alpha}); \omega_1 r, \omega)\gamma^{(2)}( 2\alpha + \omega_2(r- 2n_{\alpha});\omega_2 r, \omega) \;,  \label{mirror}
\end{align}
where $\alpha$, $\beta$ are fugacities for the global symmetries and we use the shorthand $\omega=\omega_1+\omega_2$.

\section{Pentagon identity for mirror symmetry}

For a special limit of flavor fugacities, the first duality  can be reduced to the mirror symmetry for $\mathcal N=2$ supersymmetric gauge theories. We will make the breaking of flavor symmetry on the level of partition functions. Here we follow the procedure described in \cite{Sarkissian:2022mlb}. First we take the limit $a_3 \to \infty$ and of course $b_3=\omega -a_1 - a_2 - a_3 - b_1 - b_2$ with the balancing condition $\sum_{i=1}^3 u_i + v_i = r$. If we collect terms coming from relevant Bernoulli polynomials which are not used in limit operations on the left-hand side, the following remains:
\begin{align} \nonumber
        \sum\limits_{y = 0}^{[r/2]} \epsilon(y)  \int_{-i\infty}^{i\infty} & e^{\frac{i\pi}{2} A } \prod _{i=1}^{2}\gamma^{(2)}(a_i -z +\omega_1(u_i - y); +\omega_1 r, \omega) \gamma^{(2)}(a_i - z) +\omega_2(r-(u_i - y)); \omega_2 r, \omega)\\ \nonumber
        &\times \gamma^{(2)}( b_i + z + \omega_1(v_i + y); \omega_1 r, \omega) \gamma^{(2)}(b_i + z + \omega_2(r-(v_i + y)); \omega_2 r, \omega) \frac{dz}{i r\sqrt{\omega_1\omega_2}}\\ \nonumber
        &= \prod\limits_{i,j=1}^{2} \gamma^{(2)}(a_i+ b_j +\omega_1(u_i + v_j); \omega_1 r, \omega)\gamma^{(2)}(a_i + b_j + \omega_2(r-(u_i + v_j)); \omega_2 r, \omega) \\ \nonumber
        & \qquad \quad \times \gamma^{(2)}(\omega - a_1 -b_1 -a_2-b_2 + \omega_1(r-u_1-v_1-u_2-v_2); \omega_1 r, \omega) \\
        & \qquad \quad \times \gamma^{(2)}(\omega - a_1 -b_1 -a_2-b_2 + \omega_2(u_1+v_1+u_2+v_2); \omega_2 r, \omega) ,
\end{align}
where the exponential factor is
\begin{equation*}
    \begin{split}
        A= \frac{1}{\omega_1 \omega_2 r} \big(2z(a_1+b_1+a_2+b_2) + (a_2+b_2)(-a_1+b_1)+(a_1+b_1)(-a_2+b_2)\big) \\
        + \frac{2y}{r}\big(u_1+v_1+u_2+v_2\big)- \frac{2}{\omega r}\big(\omega_1 - \omega_2 \big) \big(u_1 u_2 -v_1 v_2\big) .
    \end{split}
\end{equation*}
In the latter expression, we used the balancing condition to rewrite $u_3+v_3$ in terms of other flavor charges as $u_3+v_3=r-u_1-u_2-v_1-v_2$.

Now if we take $a_2 \to \infty$ limit, and define $a_1 + b_1 = g_1$, $a_2 +b_2=g_2$, $u_1 + v_1 = t_1$, $u_2 + v_2= t_2$ while sending\footnote{ Actually in the latter expression we only have the following combinations $u_1+v_1$ and $a_1+b_1$, therefore by sending we do not lose any information.} $b_1, v_1 \to 0$, we get:
\begin{align} \nonumber
        \sum\limits_{y = 0}^{[r/2]}  \int_{-i\infty}^{i\infty} &e^{\frac{2\pi i z}{\omega_1 \omega_2 r} (\frac{2g_1 + g_2 - \omega}{2}) + \frac{i \pi y}{r} (t_1  + 2t_2 - r)}\gamma^{(2)}(z +\omega_1 y; \omega_1 r, \omega) \gamma^{(2)}(z +\omega_2(r-y)); \omega_2 r, \omega)\\ \nonumber
        & \times \gamma^{(2)}(-z+g_1 +\omega_1(t_1 - y); \omega_1 r, \omega) \gamma^{(2)}(-z+g_1 +\omega_2(r-t_1+ y)); \omega_2 r, \omega)\frac{dz}{i r\sqrt{\omega_1\omega_2}} \\ \nonumber
        &=e^{\frac{2\pi i }{\omega_1 \omega_2 r} (\frac{g_1}{2}) (\frac{2g_2 + g_1 - \omega}{2}) + \frac{i \pi}{r} ( \frac{-t_1}{2})(t_1 + 2t_2 - r)} \gamma^{(2)}(\omega - g_1 - g_2 + \omega_1(r - t_1 -t_2); \omega_1 r, \omega)\\ \nonumber
        & \qquad \times \gamma^{(2)}(\omega - g_1 - g_2 + \omega_2(t_1 + t_2);\omega_2 r, \omega) \gamma^{(2)}(g_2 + \omega_1 t_2; \omega_1 r, \omega) \gamma^{(2)}( g_2 + \omega_2(r-t_2);\omega_2 r, \omega) \\
        & \qquad \times \gamma^{(2)}(g_1 + \omega_1 t_1; \omega_1 r, \omega)\gamma^{(2)}( g_1 + \omega_2(r- t_1);\omega_2 r, \omega)     \label{lastpentagon}
\end{align}
As one can see, the expression (\ref{lastpentagon}) has the form of the integral pentagon relation. The redefinition of the fugacities as $g_1/2 \to \alpha$, $z \to z + \alpha$, $g_2 \to \beta - \alpha $ and $t_1 \to 2n_{\alpha} $, $y \to y + n_{\alpha}$, $t_2 \to n_{\beta} - n_{\alpha}$ gives the integral identity (\ref{mirror}) for mirror symmetry \cite{Imamura:2012rq}. The case $r=1$ corresponds to the identity from \cite{Sarkissian:2022mlb,Faddeev:2000if}.

\section{Summary}

We report a solution to the pentagon equation obtained via reduction of equality for $S_b^3/\mathbb Z_r$ partition functions  of a certain $\mathcal N=2$ supersymmetric duality via the approach described in \cite{Sarkissian:2022mlb}.

There are many possible directions for future research.  The supersymmetric partition functions are related to topological invariants of triangulated three-manifolds via $3-3$ relation \cite{Dimofte:2011ju,Dimofte:2012pd}. The special  $r=1$ case of integral identities discussed here is used in \cite{Kashaev:2012cz} for the construction of the Turaev-Viro invariant of three-manifolds. It would be interesting to construct the invariant of the lens space \cite{suzuki2002turaev}. Much work remains to be done in this direction.

It would be interesting to construct a solution to the corresponding Yang-Baxter equation using the obtained solution to the pentagon equation in the context of gauge/YBE correspondence \cite{Gahramanov:2017ysd,Yamazaki:2018xbx}. 

Another possible future direction is obtaining the five-term identity \cite{Alexandrov:2015xir} implied by Kontsevich-Soibelman wall-crossing via the procedure described in \cite{Faddeev:2011tah}. 



\section*{Acknowledgements} 

It is pleasure to thank Mustafa Mullahasanoglu for the helpful discussions. The work of Ilmar Gahramanov is partially supported by TUBITAK grant 220N106, by the 1002-TUBITAK Quick Support Program under grant number 121F413, and by the Russian Science Foundation grant number 22-72-10122.

\bibliographystyle{utphys}
\bibliography{pentagon}

\end{document}